%% file: 9510090.tex
\documentstyle[aps,prl,floats,eepic,epsf]{revtex}

\def\mydate{October 13, 1995}
\def\myend{\end{document}}

\normalsize

\newcounter{sxn}

\newcounter{axn}

\newcommand{\beeq}{\begin{equation}}
\newcommand{\eneq}{\end{equation}}
\newcommand{\beqn}{\begin{eqnarray}}
\newcommand{\eeqn}{\end{eqnarray}}

\def\dd{\partial}
\def\la{\raise.16ex\hbox{$\langle$} \, }
\def\ra{\, \raise.16ex\hbox{$\rangle$} }
\def\go{\rightarrow}

\def\onehalf{ \hbox{${1\over 2}$} }
\def\onethird{ \hbox{${1\over 3}$} }
\def\psibar{ \psi \kern-.65em\raise.6em\hbox{$-$} }
\def\mbar{ m \kern-.78em\raise.4em\hbox{$-$}\lower.4em\hbox{} }

\def\Bbar{ B \kern-.73em\raise.6em\hbox{$-$}\hbox{} }

\def\vphi{ {\varphi} }

\def\tot{{\rm tot}}

\def\mass{{\rm mass}}
\def\osc{{\rm osci}}
\def\vac{{\rm vac}}

\def\LapN{{\triangle_N^\vphi}}
\def\potN{{V_N(\vphi)}}

\def\boxit#1{$\vcenter{\hrule\hbox{\vrule\kern3pt
     \vbox{\kern3pt\hbox{#1}\kern3pt}\kern3pt\vrule}\hrule}$}
\def\bigbox#1{$\vcenter{\hrule\hbox{\vrule\kern5pt
     \vbox{\kern5pt\hbox{#1}\kern5pt}\kern5pt\vrule}\hrule}$}

\begin{document}

\title{\bf{The interplay between mass, volume, $\theta$, and
$\la \psibar\psi\ra$  in $N$-flavor QED$_2$ }}
\author{J. E. Hetrick$^1$, Y. Hosotani$^2$, and S. Iso$^3$}
\address{ $^1$Department of Physics, University of Arizona, Tucson,
Arizona 85721, USA}
\address{ $^2$School of Physics and Astronomy, University of Minnesota,
Minneapolis, Minnesota 55455, USA}
\address{ $^3$Department of Physics, University of Tokyo,
Tokyo 113, Japan}
\date{\mydate}
\maketitle

\begin{abstract}
The Schwinger model (QED$_2$) with $N$ flavors of massive fermions on
a circle of circumference $L$, or equivalently at finite temperature
$T$, is reduced to a quantum mechanical system of $N-1$ degrees of
freedom.  With degenerate fermion masses ($m$) the chiral condensate
develops a cusp singularity at $\theta=\pm \pi$ in the limit $L \go
\infty$ or $T \go 0$,which is removed by a large asymmetry in the
fermion masses.  Physical quantities sensitively depend on the
parameter $mL$ or $m/T$, and the $m\go 0$ and $L\go\infty$ (or $T\go
0$) limits do not commute.  A detailed analysis is given for $N=3$.
\\
Report numbers: AZPH-TH/95-25, UMN-TH-1410/95, UT-722
\\
E-print: {\sf hep-th/9510090}

\end{abstract}


Two-dimensional QED has profound resemblance to four-dimensional QCD,
including chiral symmetry breaking, confinement, instantons, and
$\theta$ vacua \cite{SchwGEN}. Defined on a
circle\cite{Manton,HH,SchwS1}, the model is mathematically equivalent
to its finite temperature version\cite{SchwFT}, and recently the model
with general fermion content was studied in connection with $Z_n$
symmetry breaking \cite{HNZ}.  A subtle
difference between the massless and massive theories has been noted
there, for which the non-commutativity of the massless limit and zero
temperature limit is an crucial factor.  In lattice gauge theory,
the commutivity of the two limits $m\go 0$ and $L\go\infty$ is subtle
and important. This is particularly true in the issue of the
triviality of QED$_4$ at strong coupling\cite{QEDtriv}.

In a previous paper\cite{HHI} we have shown that massive
two-flavor QED$_2$ defined on a circle is reduced to the
quantum mechanics of a pendulum.
The physics is controlled by the strength of the
pendulum potential $\kappa$, which is given in a large volume $L$ by
$\kappa \sim mL(e L)^{1/2} |\cos\onehalf\theta|$ where $e$, $m$, and $\theta$
are the coupling constant, fermion mass, and vacuum angle, respectively.
It was recognized that $L\go\infty$ and $m\go 0$ limits do not commute.
Here we generalize our analysis to the $N$-flavor case.

The $N\ge 2$ Schwinger model is distinctively different from the $N=1$
model. With massless fermions the spectrum
contains $N-1$ massless bosons (mesons), and the chiral condensate
vanishes, $\la \psibar\psi\ra=0$.\cite{Halpern} With massive fermions,
the boson masses and chiral condensate are non-vanishing, but have
singular dependence on $m$ and $\theta$ at $\theta=\pm\pi$ in the
$L\go\infty$ limit.\cite{Coleman}

Suppose that the fermion and gauge fields obey anti-periodic and
periodic boundary conditions, respectively.  The model, after a Wick
rotation, is equivalent to the model on a line at finite temperature
$T=L^{-1}$.

In the bosonization method
\beeq
\psi^a_\pm = {C^a_\pm\over \sqrt{L}} \,
 e^{\pm i \{ q^a_\pm + 2\pi p^a_\pm (t \pm x)/L \} }
  : e^{\pm i\sqrt{4\pi} \phi^a_\pm (t,x) }:
\label{bosonize}
\eneq
where $C^a_\pm$ is the Klein factor. We refer the reader to \cite{HHI}
for details. Anti-Periodicity for the fermions is ensured by a physical
state condition $e^{2\pi i p^a_\pm} ~ | \, {\rm phys} \ra = |\, {\rm
phys} \ra$.  Thus $p^a_\pm$ takes integer values.

The Hamiltonian in the Schr\"odinger picture becomes
\beqn
&&H_\tot = H_0 + H_\osc  + H_\mass + ({\rm constant}) \cr
\noalign{\kern 8pt}
&&H_0 =  -{e^2 L\over 2} {d^2\over d\Theta_W^2}
 + {N\over 2\pi L} \Big\{ \Theta_W +
    {\pi\over N} \sum_{a=1}^N (p^a_++p^a_-) \Big\}^2
- {\pi\over 2NL} \Big\{ \sum_{a=1}^N (p^a_++p^a_-) \Big\}^2
+ {\pi\over L} \sum_{a=1}^N \Big\{ (p^a_+)^2
   + (p^a_-)^2 \Big\}\cr
\noalign{\kern 8pt}
&&H_\osc = \int dx \, {1\over 2}  \bigg[
  \sum_{a=1}^N \Big\{ \, \Pi_a^2 + (\phi_a')^2 \Big\}
+{e^2\over \pi}  \Big( \sum_{a=1}^N \phi_a  \Big)^2  \bigg] .
    \label{Hamiltonian}
\eeqn
$\Theta_W$ is the Wilson line phase around the circle:
$A_1=\Theta_W(t)/eL$.  It couples to $p^a_\pm$'s through the chiral
anomaly\cite{HH}.  $\phi_a=\phi^a_++\phi^a_-$ and $\Pi_a$ is its
canonical conjugate.

In the absence of fermion masses
the Hamiltonian is exactly solvable.  The spectrum  contains one massive field
$N^{-1/2}\sum_{a=1}^N \phi_a$ with a mass $\mu=(N/\pi)^{1/2} e$, and
$N-1$ massless fields.  The mass term
$H_\mass=\int_0^L dx \sum_a m_a  \psibar_a\psi_a$ however gives various
interactions among the zero and $\phi$ modes.  We present an
algorithm valid for $|m_a| \ll e$ to evaluate the effects of $H_\mass$.
We stress that this by no means implies that $H_\mass$ can be treated as a
small perturbation.  On the contrary, it dominates over $H_0$ in the
$L\go\infty$ or $T\go 0$ limit; its effects are quite non-perturbative.

As $H_\mass$ commutes with $p^a_+-p^a_-$, we restrict ourselves
to states with $(p^a_+-p^a_-) \, | \, {\rm phys} \ra =0$.  Then
a complete set of eigenfunctions of
$H_0$ is $\Phi^{(n_1, \cdots, n_N)}_s \sim
u_s\big[\Theta_W+(2\pi\sum_a n_a/ N) \big] \, e^{i\sum n_a  (q^a_++q^a_-)}$
where $u_s$ satisfies $\onehalf (-\dd_x^2 + x^2) u_s = (s+\onehalf) u_s$
with $\Theta_W= (\pi e^2 L^2/ N)^{1/4} x $.
The ground states of $H_0$ are  infinitely degenerate
for $n_1= \cdots=n_N$.

It proves to be much more convenient to work in a coherent state basis
$\Phi_s(\vphi_1,\cdots,\vphi_{N-1};\theta)$ given by
\beeq
\Phi_s(\vphi_a;\theta)
\sim \sum_{ \{ n,r_a \} }
e^{in\theta + i\sum r_a\vphi_a }
  ~ \Phi_s^{(n+r_1, \cdots,n+r_{N-1}, n)} \label{coherent}
\eneq
$H_\tot$ induces transitions among $\Phi_s(\vphi_a;\theta)$.
However the effect of transitions in the $s$ index is small
\cite{HHI} and we restrict
ourselves to $s=0$ states. The vacuum state is written as
\beeq
\Phi_\vac(\theta) = \int_0^{2\pi} d\vphi_1 \cdots d\vphi_{N-1} ~
f(\vphi_a;\theta) ~ \Phi_0(\vphi_a;\theta) ~.   \label{vacuum}
\eneq

$H_\mass$ significantly alters the vacuum structure of the
$\phi_a$ modes as well.  Its main effect is that the $N-1$ previously
massless fields now acquire finite masses.  The vacuum
is defined with respect to these physical fields.

We write mass eigenstate fields as $\chi_\alpha = U_{\alpha a}
\phi_a$.  In this physical space $(p^a_+-p^a_-) \, | \, {\rm phys} \ra
=0$, the fermion mass operator $M_a=\psibar_a\onehalf(1+\gamma^5)\psi_a$
is (in the Schr\"odinger picture)
\beqn
&&M_a =
  - {e^{i ( q^a_- + q^a_+ )} \over L}
   \prod_{\alpha=1}^N  B(\mu_\alpha L)^{{(U_{\alpha a})}^2}
  N_{\mu_\alpha} [ e^{iU_{\alpha a}\sqrt{4\pi}\chi_\alpha} ]  \cr
\noalign{\kern 5pt}
&&B(z) ={z\over 4\pi} \exp \bigg\{ \gamma + {\pi\over z}
 - 2 \int_1^\infty  {du \over (e^{uz} - 1)\sqrt{u^2-1}}  \bigg\}.
\label{massOperator}
\eeqn
Here $N_{\mu}[\cdots]$ indicates normal-ordering with reference to a mass
$\mu$.   We have made use of
$N_0[e^{i\beta\chi}]= B(\mu L)^{\beta^2/4\pi} N_{\mu}
[e^{i\beta\chi}]$.\cite{HH} ~  It is easy to find
\beqn
\la \Phi_s(\theta'; \vphi'_a) | H_\mass
 | \Phi_s(\theta; \vphi_a) \ra
&=& - \delta_{2\pi}(\theta'-\theta)
 \prod_{b=1}^{N-1}\delta_{2\pi}(\vphi'_b -\vphi_b)
\sum_{a=1}^N 2 m_a A_a \cos \vphi_a \cr
A_a &=&  e^{-\pi/N\mu L} \,
\prod_{\alpha=1}^N B(\mu_\alpha L)^{(U_{\alpha a})^2} ~~,
\label{massMatrix}
\eeqn
where $\vphi_N = \theta - \sum_{a=1}^{N-1} \vphi_a$.

The  equation $H_\tot \, \Phi_\vac(\theta) = E\,\Phi_\vac(\theta)$
becomes:~~$\displaystyle{
\Big\{ -\LapN + \potN \Big\} ~ f(\vphi)=  {NEL\over 2\pi(N-1)}~
f(\vphi)}$,~~where
\beqn
&&\LapN =
\sum_{a=1}^{N-1} {\dd^2\over \dd\vphi_a^2}
-{2\over N-1} \sum_{a<b}^{N-1} {\dd^2\over \dd\vphi_a \dd\vphi_b} \cr
\noalign{\kern 3pt}
&&\potN = -{NL\over (N-1)\pi}
{}~  \sum_{a=1}^N  m_a A_a  \cos \vphi_a ~~.
\label{QMeq}
\eeqn

The potential $\potN$
depends, through $A_\alpha$ defined in (\ref{massMatrix}), on
$\mu_\alpha$ and $U_{\alpha a}$ which are to be self-consistently
determined from the wave function $f(\vphi_a;\theta)$.  In the
$\theta$-vacuum  (\ref{vacuum}),
$\la M_a \ra_\theta = - (A_a/L) \,  \la e^{-i\vphi_a} \ra_f$
where the $f$-average is defined by
$\la g(\vphi) \ra_f = \int [d\vphi] \, g(\vphi)  |f(\vphi) |^2$.

$\potN$ has a similar structure to the potential which appears in
the effective chiral Lagrangian of QCD.
In Witten's formalism\cite{Witten} the $\vphi_a$'s are related to
the pseudoscalar mesons themselves.  The constraint $\sum_{a=1}^N
\vphi_a=\theta$ appears when the chiral anomaly dominates over
the quark masses.  In our case the $\chi_\alpha$'s represent the boson
fields, whereas $\vphi_a$'s are parameters of the coherent state
basis.  The similar structure emerges as a result of the
pattern of $SU(N)\times SU(N)$ symmetry breaking.

To find the boson masses $\mu_\alpha$,  we denote
\beeq
\pmatrix{R_a \cr I_a \cr}
= {8\pi\over L} \, m_a A_a \cdot
 \pmatrix{{\it Re}\cr {\it Im}\cr}  \la e^{-i\vphi_a} \ra_f ~.
\label{RIdef}
\eneq
$H_\mass$ yields $\prod_\alpha
N_{\mu_\alpha} [ e^{iU_{\alpha a}\sqrt{4\pi}\chi_\alpha} ] $.
Expanding $\chi_\alpha$ in $H_\mass$ and adding the contribution from
the chiral anomaly, one finds
\beeq
H^\chi_\mass = \int dx ~ \bigg\{
{U_{\alpha a} I_a \over \sqrt{4\pi}} \,  \chi_\alpha
+ {1\over 2} \, \mu_\alpha^2 \, \chi_\alpha^2
   + {\rm O}(\chi^3) \, \bigg\} ~.
\label{chiMass}
\eneq
Here $U_{\alpha a}$ diagonalizes the matrix
\beqn
&& {\mu^2\over N} \pmatrix{ 1 &\cdots & 1 \cr
                                     \vdots & \ddots & \vdots\cr
                                     1 & \cdots & 1 \cr}
  + \pmatrix{ R_1 &&\cr
              &\ddots&\cr
              &&R_N\cr}~~,
\label{diagonalizeMass}
\eeqn
and $\mu_\alpha$ are the eigenvalues.
(\ref{massMatrix}), (\ref{QMeq}),  (\ref{RIdef}), and
the diagonalization of (\ref{diagonalizeMass}) must be solved simultaneously.

Suppose that all fermion masses are degenerate: $m_a =
m \ll e$.  In this case  $R_a = R$ and $\la e^{-i\vphi_a}
\ra_f = \la e^{-i\vphi} \ra_f$.  One can choose $U_{\alpha a}$ such
that $U_{1a} = N^{-1/2}$, $\mu_1^2 = \mu^2 + R$,  and
$\mu_2^2 = \cdots = \mu_N^2 = R$.
The potential and boson masses are reduced to
\beqn
&&\potN = ~-  \kappa_0 ~  \sum_{a=1}^N \cos \vphi_{a} \cr
\noalign{\kern 4pt}
&&\kappa_0 = {NmL\over (N-1)\pi}  ~B(\mu_1 L)^{{1\over N}}
B(\mu_2 L)^{1-{1\over N}} ~ e^{-\pi/N\mu L} \cr
\noalign{\kern 6pt}
&&{\mu_2^2\over 4\pi}  = {2\pi \over L^2} \, {N-1\over N} \, \kappa_0 \,
    \la \cos \vphi \ra_f
= -m\, \la \psibar_a \psi_a \ra_\theta ~~.
   \label{potential2}
\eeqn
The last relation is analogous to the PCAC relation in
QCD\cite{Smilga}.  The strength $\kappa_0$ and $\theta$ fix the
potential $\potN$, and thus control the physical behavior.  Since
$B(z) \sim e^\gamma z /4\pi$ for $z \gg 1$ and $B(0)=1$, $\kappa_0 \go
\infty$ (0) as $L \go \infty$ (0). So long as $m\not= 0$, the
location of the minimum of the potential determines the physics
at $L\go\infty$.

The potential $\potN$ has a minimum at
\beeq
\vphi_1 = \cdots = \vphi_N = {1\over N} \Big( \theta -
    2\pi \Big[ {\theta+\pi\over 2\pi} \Big] \Big)
\equiv {1\over N}\, \bar\theta(\theta)
\label{minimum1}
\eneq
[$-\pi \le \bar\theta <\pi$];
its location jumps discontinuously  at $\theta=\pi$ from $\vphi_a=\pi/N$
to $\vphi_a=-\pi/N$.
There is a special feature in the $N=2$ case:  as $V_2(\vphi) = -2\kappa_0
\cos\onehalf\theta \cos(\vphi_1-\onehalf\theta)$, the potential vanishes
at $\theta=\pm\pi$.  Its behavior is controlled by the single parameter
$\kappa_0 |\cos\onehalf\theta|$ (see \cite{HHI}).

When $\kappa_0 \gg 1$ (or as $L\go\infty$), the wave function
$f(\vphi)$ approaches a delta function around the minimum of the potential
$\potN$. Hence $\lim_{L\go\infty} \la \cos\vphi_a\ra_f = \cos(\bar\theta/N)$.
For $m/\mu \ll 1$  one finds from (\ref{potential2})
\beeq
{1\over \mu} \, \la \psibar\psi \ra_\theta
= - {1\over 4\pi} \Big( 2e^\gamma \cos {\bar\theta\over N} \Big)^{{2N\over
N+1}}  \, \Big({m\over \mu}\Big)^{{N-1\over N+1}} ~.
\label{massInfinity}
\eneq

In the opposite limit $\kappa_0 \ll 1$, the wave function is  given
by
$f = (2\pi)^{-(N-1)/2}  \big\{ 1 + \kappa_0 \sum_{a=1}^N \cos \vphi_a
+ \cdots \big\}$
so that
\beeq
\la \cos \vphi_a \ra_f
=\cases{(1+\cos\theta) \kappa_0  &for $N=2$\cr
        \kappa_0  &for $N\ge 3$.\cr}   \label{smallKappa}
\eneq
For $N \ge 3$ and  $\mu L \ll 1$,
\beeq
{1\over \mu}\, \la \psibar\psi \ra_\theta = - {2N\over \pi(N-1)} \,
 { m\over \mu} \,  e^{-2\pi/N\mu L}.
       \label{smallMass}
\eneq
In the intermediate region, $\mu L \gg 1 \gg \mu_2 L$,
\beeq
{1\over \mu} \, \la \psibar\psi \ra_\theta = - {2N\over \pi(N-1)} \,
{m\over \mu} \, \Big( {\mu L\over 4\pi} \, e^\gamma \Big)^{2/N}  ~.
  \label{mediumMass}
\eneq
For $N=2$,
(\ref{smallMass}) and (\ref{mediumMass}) must be multiplied by a factor
$2 \cos^2 \onehalf\theta $.

For general values of $\mu L$ and $m/\mu$, we have determined
$\mu_2$ and $\la \psibar\psi \ra_\theta$ in the $N=3$ case
numerically.  Fig.\ 1 shows typical wave functions
$|f(\vphi)|^2$ at $\kappa_0= 0.1, 10$ and $\theta=0$, $0.999\pi$. In
fig.\ 2, $\la \psibar\psi \ra / \mu$ at $\theta=0$ and $\mu L=10^3$ is
plotted as a function of $m/\mu$.  In fig.\ 3, $\la \psibar\psi \ra /
\mu$ at $\theta=0$ is plotted as a function of $T/\mu$ and $m/\mu$.
In fig.\ 4, the $\theta$ dependence of $\la \psibar\psi
\ra / \mu$ at $m/\mu=10^{-3}$ is depicted for various $\mu L$.

Several important observations follow.  As is evident from
(\ref{massInfinity}), $\la \psibar\psi \ra$ in the infinite volume
limit has fractional power dependence on the fermion mass $m$.
However, if the massless limit $m\go 0$ is taken with a finite $L$,
then $\kappa_0$ becomes very small ($\kappa_0 \ll 1$) and $\la
\psibar\psi \ra$ is given by either (\ref{smallMass}) or
(\ref{mediumMass}), which is linear in $m$.  In this limit the fermion
mass term in the Hamiltonian can be treated as a small perturbation.
On the other hand, the effect of finite fermion masses in infinite
volume is always non-perturbative. The massless and infinite volume
limits do not commute with each other.  The important parameter is
$\kappa_0$.  We have juxtaposed a plot for $\kappa_0$ in fig.\ 2
{}from which one can see that the crossover in physical behavior
takes place around $\kappa_0=0.2$.

Our result can be reinterpreted for the Schwinger model on a line at
finite temperature by replacing $L$ by $T^{-1}$. We see that there is
no phase transition at finite temperature.\cite{SchwFT}  The condensate $\la
\psibar\psi \ra$, which is non-vanishing at $T=0$, smoothly goes to
zero at finite temperature.  See fig.\ 3.

Thirdly all physical quantities are periodic in $\theta$ with period
$2\pi$.  At $L=\infty$, $\la \psibar\psi \ra_\theta$ has a cusp at
$\theta=\pm\pi$,\cite{Coleman}  while at any finite $L$ the cusp disappears
as shown in fig.\ 4. The appearance of the cusp is traced back to the
discontinuous jump in the location of the minimum of the potential
$\potN$.

So far we have concentrated on cases with a symmetric fermion mass
term.  For a general fermion masses, the evaluation
procedure is more involved. One important conclusion can be drawn.
In the potential $\potN$ in (\ref{QMeq}), the coefficients of $\cos\vphi_a$ are
all different in general.  The potential for $N=3$ is proportional to
\beeq
F(\vphi) = -  q \cos\vphi_1 - r \cos\vphi_2
   - \cos(\theta - \vphi_1 -\vphi_2)
\eneq
where $q=m_1A_1/ m_3A_3$ and $r=m_2A_2/ m_3A_3$.
We have investigated the location of the
minimum of $F(\vphi)$ as a function of $\theta$ with various parameter
values $(q,r)$.  In the symmetric case $(q,r)=(1,1)$, the
location of the minimum moves from $(\vphi_1,\vphi_2)
=(-\onethird\pi,-\onethird\pi)$ to $(\onethird\pi,\onethird\pi)$ as
$\theta$ varies from $-\pi$ to $+\pi$, and jumps to return to the
original point $(-\onethird\pi,-\onethird\pi)$; see fig.\ 5.  Now we
add an asymmetry.  Several cases are plotted in fig.\ 5.  One can see
that so long as the asymmetry is small enough, there is a
discontinuous jump at $\theta=\pm\pi$. However, a sufficiently large
mass asymmetry removes this discontinuity.  For instance, for $(q,r) =
(1, 0.3)$, the minimum moves from $(\vphi_1,\vphi_2)=(0,-\pi)$ to
$(0,+\pi)$, hence making a closed loop in the $\vphi_1$--$\vphi_2$
plane.  We have observed that with sufficiently large asymmetry, the
minimum at $\theta=\pm\pi$ is located at either $(0,0)$, $(0,\pm\pi)$
or $(\pm\pi,0)$.  These three points are related by $S_3$
transformations.

We conclude that a sufficiently large asymmetry in the fermion masses
removes the cusp singularity at $\theta=\pm\pi$ in
$\la \psibar\psi \ra$ in the $L\go\infty$ limit.  A similar conclusion
has been drawn in the QCD context by Creutz.\cite{Creutz}

\bigskip
\leftline{\bf Acknowledgments}

This work was supported in part by by the U.S.\ Department of Energy
under contracts DE-FG03-95ER-40906 (J.H.) and by DE-AC02-83ER-40105
(Y.H.).  J.H. would like to thank the ITP at U.C. Santa Barbara and
the TPI at U. Minnesota, and Y.H. would like to thank the Aspen Center
for Physics, for their hospitality where a part of this work was
carried out.


\def\ap {{\it Ann.\ Phys.\ (N.Y.)} }
\def\cmp {{\it Comm.\ Math.\ Phys.} }
\def\ijmpA {{\it Int.\ J.\ Mod.\ Phys.} { \bf A}}
\def\ijmpB {{\it Int.\ J.\ Mod.\ Phys.} { \bf B}}
\def\jmp {{\it  J.\ Math.\ Phys.} }
\def\mplA {{\it Mod.\ Phys.\ Lett.} { \bf A}}
\def\mplB {{\it Mod.\ Phys.\ Lett.} { \bf B}}
\def\plB {{\it Phys.\ Lett.} { \bf B}}
\def\plA {{\it Phys.\ Lett.} { \bf A}}
\def\nc {{\it Nuovo Cimento} }
\def\npB {{\it Nucl.\ Phys.} { \bf B}}
\def\pr {{\it Phys.\ Rev.} }
\def\prl {{\it Phys.\ Rev.\ Lett.} }
\def\prB {{\it Phys.\ Rev.} { \bf B}}
\def\prD {{\it Phys.\ Rev.} { \bf D}}
\def\prp {{\it Phys.\ Report} }
\def\ptp {{\it Prog.\ Theoret.\ Phys.} }
\def\rmp {{\it Rev.\ Mod.\ Phys.} }
\def\hep {{\tt hep-th/}}


\rule[3mm]{15cm}{1mm} \\
\leftline{\bf Figures:}
\vskip 4cm

\begin{figure}[htb]
\epsfxsize= 15cm
\epsffile[50 250 525 560]{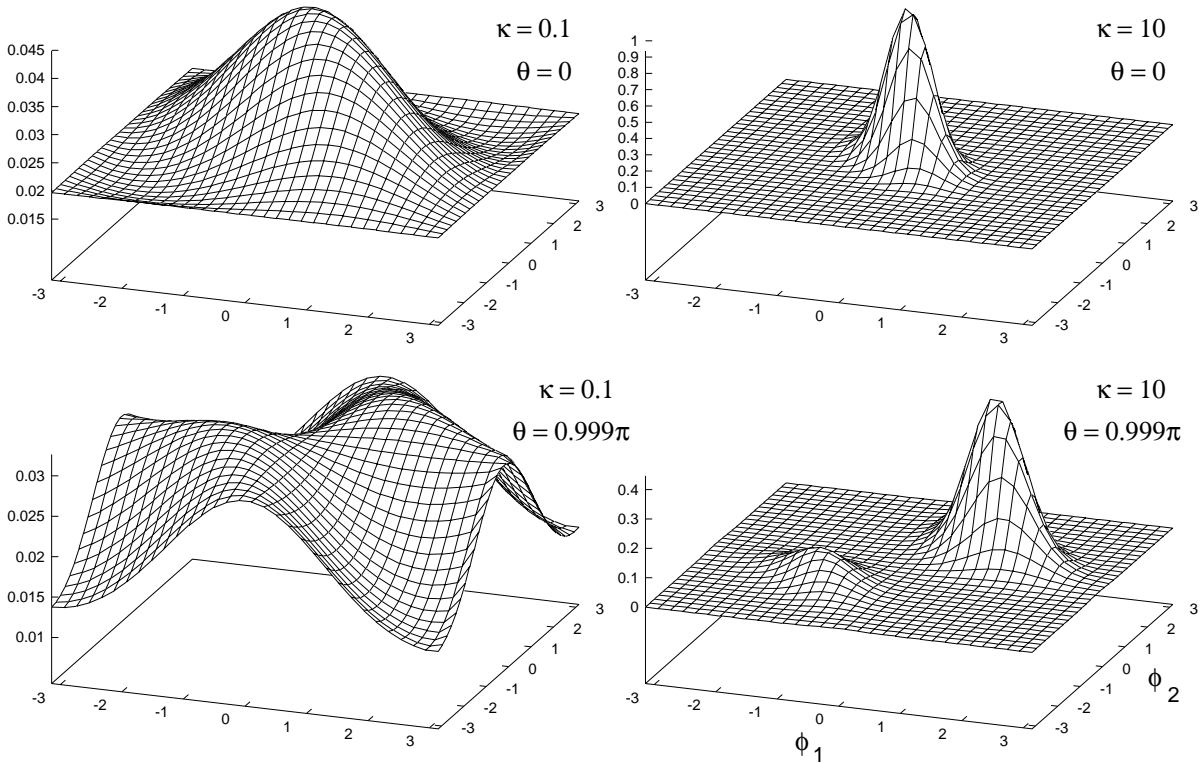}
\vskip 1cm
\caption{Typical wave functions $|f(\vphi)|^2$ at $\kappa_0 = 0.1$ and
10, and $\theta = 0$ and $0.999\pi$}
\label{fig:1}
\vskip 0.2cm
\end{figure}

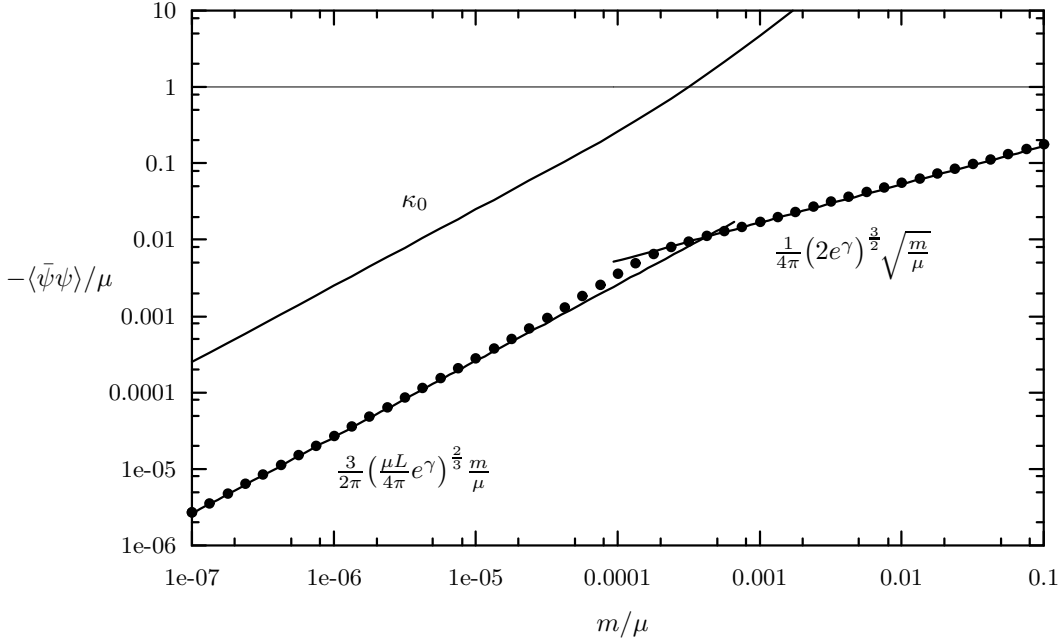
\begin{figure}[htb]
  \leftline{
       \input{PBPprep.tex}  }
\vskip 1cm
\caption{The behavior of $=-\la \psibar \psi \ra/\mu$ versus
$m/\mu$ at $\mu L=10^3$ and $\theta=0$. The analytic forms, (13) and (16),
 are also displayed. The crossover occurs at  $\kappa_0 \approx 0.2$.}
\label{fig:2}
\vskip 0.2cm
\end{figure}

\begin{figure}[htb]
\epsfxsize= 15cm
\epsffile[100 225 500 460]{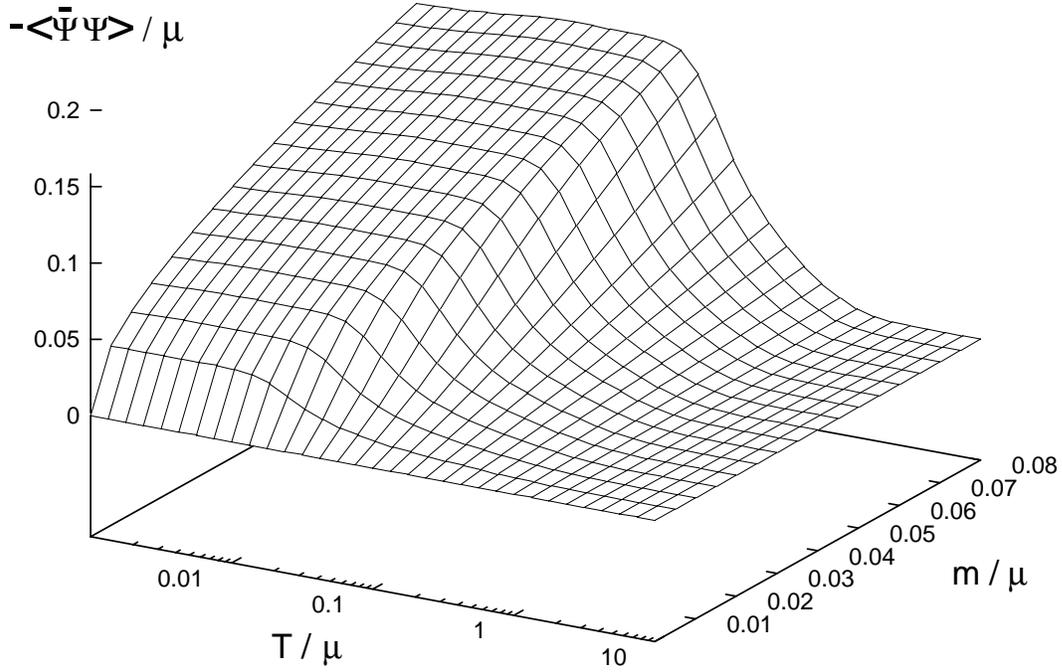}
\vskip 1cm
\caption{The chiral condensate $\la \psibar \psi \ra/\mu$ as a
function of temperature $T/\mu$ (or $1/\mu L$) at $\theta=0$.}
\label{fig:3}
\vskip 0.2cm
\end{figure}

\begin{figure}[htb]
\epsfxsize= 15cm
\epsfysize= 9cm
\epsffile[50 155 540 500]{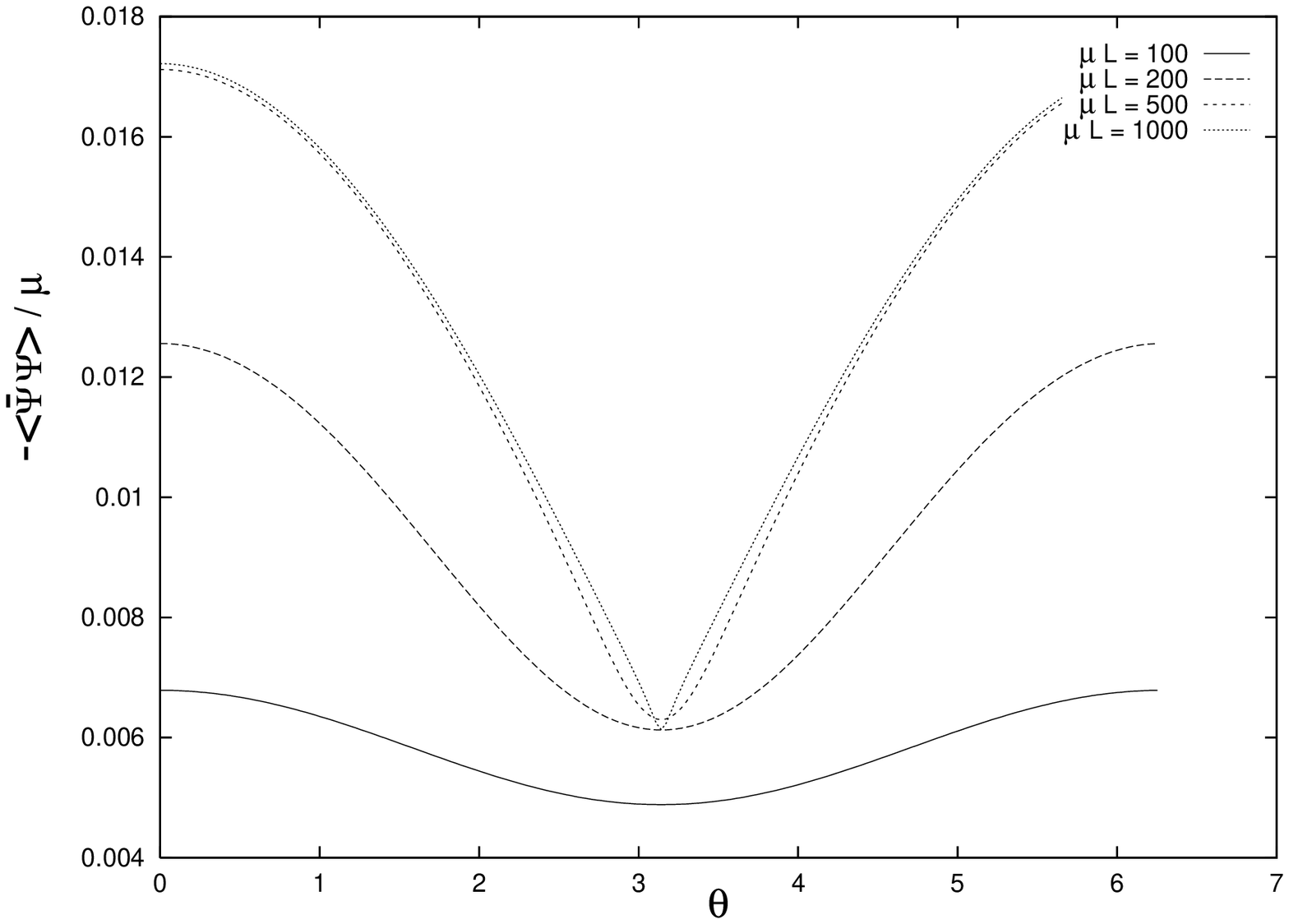}
\vskip 1cm
\caption{The $\theta$ dependence of $\la \psibar \psi\ra$. In each
case $m/\mu = 0.001$.}
\label{fig:4}
\vskip 0.2cm
\end{figure}

\begin{figure}[htb]
\epsfxsize= 15cm
\epsfysize= 9cm
\epsffile[60 190 540 550]{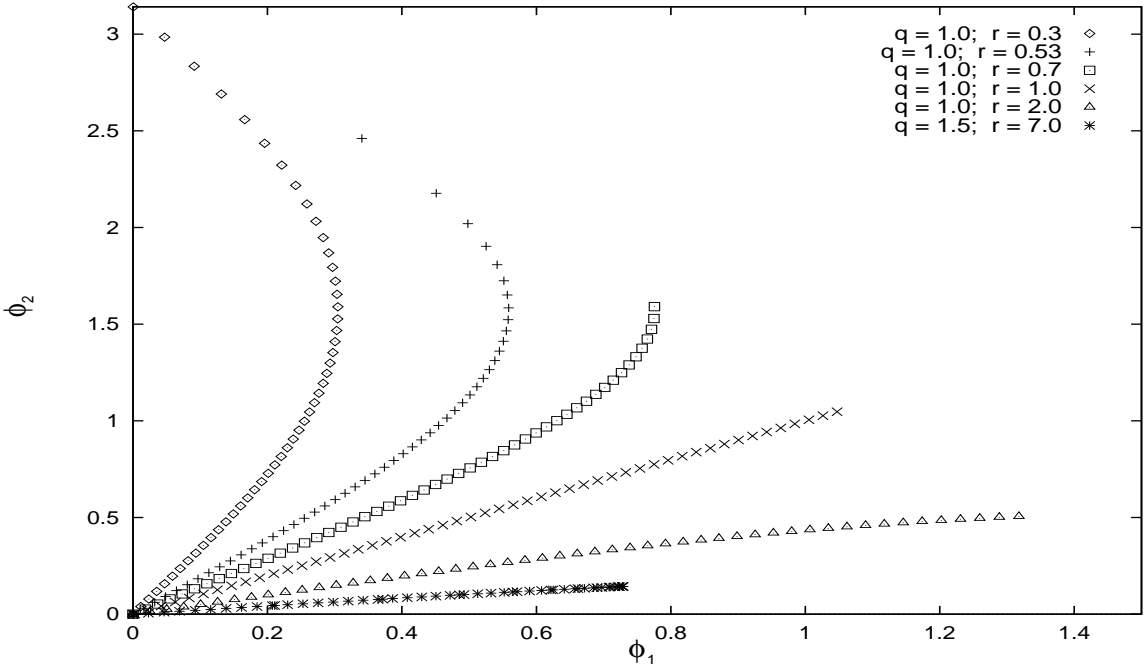}
\vskip 1cm
\caption{The location of the minimum of
$F(\vphi_1,\vphi_2)$ in (17). The points in the graph for each $q$ and
$r$ pair run from $\theta = 0$ (at the origin) and end at $\theta = \pi$.
For $\theta<0$, the minimum is located at $(\vphi_1,\vphi_2)_\theta
=-(\vphi_1,\vphi_2)_{-\theta}$.  For $(q,r)=(1.5,7)$ the minimum returns
to the origin at $\theta=\pi$.}
\label{fig:5}
\vskip 0.2cm
\end{figure}

\end{document}

%% file: PBPprep.tex
\setlength{\unitlength}{0.26500pt}
\begin{picture}(1500,900)(0,0)
\thicklines \path(220,113)(240,113)
\thicklines \path(1436,113)(1416,113)
\put(198,113){\makebox(0,0)[r]{\small 1e-06}}
\thicklines \path(220,146)(230,146)
\thicklines \path(1436,146)(1426,146)
\put(198,146){\makebox(0,0)[r]{}}
\thicklines \path(220,189)(230,189)
\thicklines \path(1436,189)(1426,189)
\put(198,189){\makebox(0,0)[r]{}}
\thicklines \path(220,212)(230,212)
\thicklines \path(1436,212)(1426,212)
\put(198,212){\makebox(0,0)[r]{}}
\thicklines \path(220,222)(240,222)
\thicklines \path(1436,222)(1416,222)
\put(198,222){\makebox(0,0)[r]{\small 1e-05}}
\thicklines \path(220,255)(230,255)
\thicklines \path(1436,255)(1426,255)
\put(198,255){\makebox(0,0)[r]{}}
\thicklines \path(220,298)(230,298)
\thicklines \path(1436,298)(1426,298)
\put(198,298){\makebox(0,0)[r]{}}
\thicklines \path(220,321)(230,321)
\thicklines \path(1436,321)(1426,321)
\put(198,321){\makebox(0,0)[r]{}}
\thicklines \path(220,331)(240,331)
\thicklines \path(1436,331)(1416,331)
\put(198,331){\makebox(0,0)[r]{\small 0.0001}}
\thicklines \path(220,364)(230,364)
\thicklines \path(1436,364)(1426,364)
\put(198,364){\makebox(0,0)[r]{}}
\thicklines \path(220,408)(230,408)
\thicklines \path(1436,408)(1426,408)
\put(198,408){\makebox(0,0)[r]{}}
\thicklines \path(220,430)(230,430)
\thicklines \path(1436,430)(1426,430)
\put(198,430){\makebox(0,0)[r]{}}
\thicklines \path(220,440)(240,440)
\thicklines \path(1436,440)(1416,440)
\put(198,440){\makebox(0,0)[r]{\small 0.001}}
\thicklines \path(220,473)(230,473)
\thicklines \path(1436,473)(1426,473)
\put(198,473){\makebox(0,0)[r]{}}
\thicklines \path(220,517)(230,517)
\thicklines \path(1436,517)(1426,517)
\put(198,517){\makebox(0,0)[r]{}}
\thicklines \path(220,539)(230,539)
\thicklines \path(1436,539)(1426,539)
\put(198,539){\makebox(0,0)[r]{}}
\thicklines \path(220,550)(240,550)
\thicklines \path(1436,550)(1416,550)
\put(198,550){\makebox(0,0)[r]{\small 0.01}}
\thicklines \path(220,582)(230,582)
\thicklines \path(1436,582)(1426,582)
\put(198,582){\makebox(0,0)[r]{}}
\thicklines \path(220,626)(230,626)
\thicklines \path(1436,626)(1426,626)
\put(198,626){\makebox(0,0)[r]{}}
\thicklines \path(220,648)(230,648)
\thicklines \path(1436,648)(1426,648)
\put(198,648){\makebox(0,0)[r]{}}
\thicklines \path(220,659)(240,659)
\thicklines \path(1436,659)(1416,659)
\put(198,659){\makebox(0,0)[r]{\small 0.1}}
\thicklines \path(220,692)(230,692)
\thicklines \path(1436,692)(1426,692)
\put(198,692){\makebox(0,0)[r]{}}
\thicklines \path(220,735)(230,735)
\thicklines \path(1436,735)(1426,735)
\put(198,735){\makebox(0,0)[r]{}}
\thicklines \path(220,757)(230,757)
\thicklines \path(1436,757)(1426,757)
\put(198,757){\makebox(0,0)[r]{}}
\thicklines \path(220,768)(240,768)
\thicklines \path(1436,768)(1416,768)
\put(198,768){\makebox(0,0)[r]{\small 1}}
\thicklines \path(220,801)(230,801)
\thicklines \path(1436,801)(1426,801)
\put(198,801){\makebox(0,0)[r]{}}
\thicklines \path(220,844)(230,844)
\thicklines \path(1436,844)(1426,844)
\put(198,844){\makebox(0,0)[r]{}}
\thicklines \path(220,866)(230,866)
\thicklines \path(1436,866)(1426,866)
\put(198,866){\makebox(0,0)[r]{}}
\thicklines \path(220,877)(240,877)
\thicklines \path(1436,877)(1416,877)
\put(198,877){\makebox(0,0)[r]{\small 10}}
\thicklines \path(220,113)(220,133)
\thicklines \path(220,877)(220,857)
\put(220,68){\makebox(0,0){\small 1e-07}}
\thicklines \path(281,113)(281,123)
\thicklines \path(281,877)(281,867)
\put(281,68){\makebox(0,0){}}
\thicklines \path(362,113)(362,123)
\thicklines \path(362,877)(362,867)
\put(362,68){\makebox(0,0){}}
\thicklines \path(403,113)(403,123)
\thicklines \path(403,877)(403,867)
\put(403,68){\makebox(0,0){}}
\thicklines \path(423,113)(423,133)
\thicklines \path(423,877)(423,857)
\put(423,68){\makebox(0,0){\small 1e-06}}
\thicklines \path(484,113)(484,123)
\thicklines \path(484,877)(484,867)
\put(484,68){\makebox(0,0){}}
\thicklines \path(564,113)(564,123)
\thicklines \path(564,877)(564,867)
\put(564,68){\makebox(0,0){}}
\thicklines \path(606,113)(606,123)
\thicklines \path(606,877)(606,867)
\put(606,68){\makebox(0,0){}}
\thicklines \path(625,113)(625,133)
\thicklines \path(625,877)(625,857)
\put(625,68){\makebox(0,0){\small 1e-05}}
\thicklines \path(686,113)(686,123)
\thicklines \path(686,877)(686,867)
\put(686,68){\makebox(0,0){}}
\thicklines \path(767,113)(767,123)
\thicklines \path(767,877)(767,867)
\put(767,68){\makebox(0,0){}}
\thicklines \path(808,113)(808,123)
\thicklines \path(808,877)(808,867)
\put(808,68){\makebox(0,0){}}
\thicklines \path(828,113)(828,133)
\thicklines \path(828,877)(828,857)
\put(828,68){\makebox(0,0){\small 0.0001}}
\thicklines \path(889,113)(889,123)
\thicklines \path(889,877)(889,867)
\put(889,68){\makebox(0,0){}}
\thicklines \path(970,113)(970,123)
\thicklines \path(970,877)(970,867)
\put(970,68){\makebox(0,0){}}
\thicklines \path(1011,113)(1011,123)
\thicklines \path(1011,877)(1011,867)
\put(1011,68){\makebox(0,0){}}
\thicklines \path(1031,113)(1031,133)
\thicklines \path(1031,877)(1031,857)
\put(1031,68){\makebox(0,0){\small 0.001}}
\thicklines \path(1092,113)(1092,123)
\thicklines \path(1092,877)(1092,867)
\put(1092,68){\makebox(0,0){}}
\thicklines \path(1172,113)(1172,123)
\thicklines \path(1172,877)(1172,867)
\put(1172,68){\makebox(0,0){}}
\thicklines \path(1214,113)(1214,123)
\thicklines \path(1214,877)(1214,867)
\put(1214,68){\makebox(0,0){}}
\thicklines \path(1233,113)(1233,133)
\thicklines \path(1233,877)(1233,857)
\put(1233,68){\makebox(0,0){\small 0.01}}
\thicklines \path(1294,113)(1294,123)
\thicklines \path(1294,877)(1294,867)
\put(1294,68){\makebox(0,0){}}
\thicklines \path(1375,113)(1375,123)
\thicklines \path(1375,877)(1375,867)
\put(1375,68){\makebox(0,0){}}
\thicklines \path(1416,113)(1416,123)
\thicklines \path(1416,877)(1416,867)
\put(1416,68){\makebox(0,0){}}
\thicklines \path(1436,113)(1436,133)
\thicklines \path(1436,877)(1436,857)
\put(1436,68){\makebox(0,0){\small 0.1}}
\thicklines \path(220,113)(1436,113)(1436,877)(220,877)(220,113)
\put(30,495){\makebox(0,0)[l]{\shortstack{\kern-2em\hbox{$-\langle\bar\psi\psi\rangle/\mu$}}}}
\put(828,23){\makebox(0,0){{ \lower2em\hbox{$m/\mu$}}}}
\put(1050,530){\makebox(0,0)[l]{${1\over 4\pi}\big(2e^\gamma\big)^{3\over 2}$
\kern-0.5em\hbox{$\sqrt{m\over \mu}$}}}
\put(425,222){\makebox(0,0)[l]{${3\over 2\pi}\big({\mu L\over 4\pi}e^\gamma\big)^{2\over 3}{m\over \mu}$}}
\put(519,602){\makebox(0,0)[l]{$\kappa_0$}}
\put(220,160){\circle*{12}}
\put(220,160){\circle*{12}}
\put(245,173){\circle*{12}}
\put(271,187){\circle*{12}}
\put(296,201){\circle*{12}}
\put(321,214){\circle*{12}}
\put(347,228){\circle*{12}}
\put(372,242){\circle*{12}}
\put(397,255){\circle*{12}}
\put(423,269){\circle*{12}}
\put(448,283){\circle*{12}}
\put(473,297){\circle*{12}}
\put(499,310){\circle*{12}}
\put(524,324){\circle*{12}}
\put(549,338){\circle*{12}}
\put(575,352){\circle*{12}}
\put(600,366){\circle*{12}}
\put(625,380){\circle*{12}}
\put(651,394){\circle*{12}}
\put(676,408){\circle*{12}}
\put(701,423){\circle*{12}}
\put(727,438){\circle*{12}}
\put(752,453){\circle*{12}}
\put(777,469){\circle*{12}}
\put(803,485){\circle*{12}}
\put(828,501){\circle*{12}}
\put(853,516){\circle*{12}}
\put(879,529){\circle*{12}}
\put(904,539){\circle*{12}}
\put(929,547){\circle*{12}}
\put(955,555){\circle*{12}}
\put(980,562){\circle*{12}}
\put(1005,568){\circle*{12}}
\put(1031,575){\circle*{12}}
\put(1056,582){\circle*{12}}
\put(1081,589){\circle*{12}}
\put(1107,597){\circle*{12}}
\put(1132,604){\circle*{12}}
\put(1157,611){\circle*{12}}
\put(1183,618){\circle*{12}}
\put(1208,624){\circle*{12}}
\put(1233,631){\circle*{12}}
\put(1259,637){\circle*{12}}
\put(1284,644){\circle*{12}}
\put(1309,651){\circle*{12}}
\put(1335,658){\circle*{12}}
\put(1360,664){\circle*{12}}
\put(1385,672){\circle*{12}}
\put(1411,679){\circle*{12}}
\put(1436,686){\circle*{12}}
\thicklines \path(220,158)(220,158)(232,165)(245,172)(257,178)(269,185)(281,191)(294,198)(306,205)(318,211)(331,218)(343,224)(355,231)(367,238)(380,244)(392,251)(404,258)(417,264)(429,271)(441,277)(453,284)(466,291)(478,297)(490,304)(503,310)(515,317)(527,324)(539,330)(552,337)(564,344)(576,350)(588,357)(601,363)(613,370)(625,377)(638,383)(650,390)(662,396)(674,403)(687,410)(699,416)(711,423)(724,429)(736,436)(748,443)(760,449)(773,456)(785,463)(797,469)(810,476)(822,482)(822,482)(834,489)(846,496)(859,502)(871,509)(883,515)(896,522)(908,529)(920,535)(932,542)(945,549)(957,555)(969,562)(982,568)(994,575)
\thicklines \path(822,518)(834,521)(846,524)(859,528)(871,531)(883,534)(896,538)(908,541)(920,544)(932,548)(945,551)(957,554)(969,558)(982,561)(994,564)(1006,567)(1018,571)(1031,574)(1043,577)(1055,581)(1068,584)(1080,587)(1092,591)(1104,594)(1117,597)(1129,601)(1141,604)(1153,607)(1166,610)(1178,614)(1190,617)(1203,620)(1215,624)(1227,627)(1239,630)(1252,634)(1264,637)(1276,640)(1289,644)(1301,647)(1313,650)(1325,653)(1338,657)(1350,660)(1362,663)(1375,667)(1387,670)(1399,673)(1411,677)(1424,680)(1436,683)
\thicklines \path(220,375)(220,375)(220,375)(245,388)(271,402)(296,415)(321,429)(347,443)(372,456)(397,470)(423,484)(448,497)(473,511)(499,525)(524,538)(549,552)(575,566)(600,579)(625,593)(651,606)(676,620)(701,634)(727,648)(752,661)(777,675)(803,689)(828,704)(853,719)(879,735)(904,751)(929,768)(955,786)(980,804)(1005,822)(1031,841)(1056,860)(1078,877)
\thinlines \path(220,768)(220,768)(232,768)(245,768)(257,768)(269,768)(281,768)(294,768)(306,768)(318,768)(331,768)(343,768)(355,768)(367,768)(380,768)(392,768)(404,768)(417,768)(429,768)(441,768)(453,768)(466,768)(478,768)(490,768)(503,768)(515,768)(527,768)(539,768)(552,768)(564,768)(576,768)(588,768)(601,768)(613,768)(625,768)(638,768)(650,768)(662,768)(674,768)(687,768)(699,768)(711,768)(724,768)(736,768)(748,768)(760,768)(773,768)(785,768)(797,768)(810,768)(822,768)
\thinlines \path(822,768)(834,768)(846,768)(859,768)(871,768)(883,768)(896,768)(908,768)(920,768)(932,768)(945,768)(957,768)(969,768)(982,768)(994,768)(1006,768)(1018,768)(1031,768)(1043,768)(1055,768)(1068,768)(1080,768)(1092,768)(1104,768)(1117,768)(1129,768)(1141,768)(1153,768)(1166,768)(1178,768)(1190,768)(1203,768)(1215,768)(1227,768)(1239,768)(1252,768)(1264,768)(1276,768)(1289,768)(1301,768)(1313,768)(1325,768)(1338,768)(1350,768)(1362,768)(1375,768)(1387,768)(1399,768)(1411,768)(1424,768)(1436,768)
\end{picture}

%% file: 9510090.bbl
\begin{thebibliography}{99}


\bibitem{SchwGEN}
 J. Schwinger, \pr {\bf 125}  397 (1962);  {\bf 128} 2425  (1962);
J.H. Lowenstein and J.A. Swieca, \ap {\bf 68} 172  (1971).
A. Casher, J. Kogut and L. Susskind, \prD {\bf 10} 732  (1974).
S. Coleman, R. Jackiw, and L. Susskind,  \ap {\bf 93} 267  (1975).
\bibitem{Manton}  N. Manton, \ap {\bf 159} 220  (1985);
\bibitem{HH} J.E. Hetrick and Y. Hosotani, \prD {\bf 38} 2621  (1988);
\bibitem{SchwS1}
 R. Link, \prD {\bf 42} 2103  (1990).
 S. Iso and H. Murayama, \ptp {\bf 84}142  (1990) .
\bibitem{SchwFT}
 I. Sachs and A. Wipf, {\it Helv. Phys. Acta.} {\bf 65} 652 (1992).
\bibitem{HNZ}  T.H. Hansson, H.B. Nielsen and I. Zahed,
USITP-94-09, {\tt hep-ph/9405324}
\bibitem{QEDtriv} M. G\"ockeler et al, \npB {\bf 371} 713 (1992);
A. Koci\'c, J.B. Kogut and M.-P. Lombardo, \npB {\bf 398} 376 (1993);
V. Azcoiti et al.\, {\tt hep-lat/9509037}.
\bibitem{HHI}  J.E.\ Hetrick, Y.\ Hosotani and S.\ Iso, \plB {\bf 350} 92
(1995).
\bibitem{Halpern} M.B. Halpern, \prD {\bf 13} 337  (1976);
I. Affleck, \npB {\bf 265} [FS15]  448 (1986).
C. Gattringer and E. Seiler, \ap {\bf 233} 97  (1994).
H. Joos and S.I. Azakov, {\it Helv. Phys. Acta.} {\bf 67}
 723 (1994).
 M.A. Shifman and A.V. Smilga, \prD {\bf 50} 7659  (1994).
\bibitem{Coleman} S. Coleman,  \ap {\bf 101} 239  (1976).
\bibitem{Witten} E.\ Witten, \ap {\bf 128} 363 (1980).
\bibitem{Smilga} A.V. Smilga, \plB{\bf 278}  371 (1992).
\bibitem{Creutz} M.\ Creutz, BNL-61796, {\tt hep-th/9505112}.

\end{thebibliography}
